\documentclass[14 pt]{article}
\usepackage{graphicx}
\usepackage{placeins}
\usepackage{float}
\usepackage{subfigure}
\usepackage{makeidx}
\usepackage{psfig}
\usepackage{epsf}

\usepackage{times}
\usepackage{amsmath}
\usepackage{cite}

\newcommand{\comment}[1]{}
\newcommand{\ignore}[1]{}
\newcommand{\longversion}[1]{}

\def\EQ{\begin{equation}}
\def\EN{\end{equation}}

\def\RR{\hbox{I\kern-.2em\hbox{R}}}

\newcommand{\rem}[1]{}

\begin{document}

\title{\textbf{\huge {\em }{\textbf{\huge }}\textbf{\huge{} }}\\
\textbf{\huge{} }\textbf{\Large Laplacian Growth Without Surface
Tension in Filtration Combustion:  Analytical Pole Solution}}

\author{Oleg Kupervasser\\
 TRANSIST VIDEO LLC, 119296, Skolkovo Resident, Russia\\
 Department of Chemical Physics, The Weizmann Institute of Science\\
Rehovot 76100, Israel\\
 olegkup@yahoo.com;}
\maketitle
\begin{abstract}
Filtration combustion is described by Laplacian growth without surface
tension. These equations have elegant analytical solutions that replace
the complex integro-differential motion equations by simple differential
equations of pole motion in a complex plane. The main problem with
such a solution is the existence of finite time singularities. To
prevent such singularities, nonzero surface tension is usually used.
However, nonzero surface tension does not exist in filtration combustion,
and this destroys the analytical solutions. However, a more elegant
approach exists for solving the problem. First, we can introduce a
small amount of pole noise to the system. Second, for regularisation
of the problem, we throw out all new poles that can produce a finite
time singularity. It can be strictly proved that the asymptotic solution
for such a system is a single finger. Moreover, the qualitative consideration
demonstrates that a finger with $\frac{1}{2}$ of the channel width
is statistically stable. Therefore, all properties of such a solution
are exactly the same as those of the solution with nonzero surface
tension under numerical noise. The solution of the ST problem without
surface tension is similar to the solution for the equation of cellular
flames in the case of the combustion of gas mixtures.
\end{abstract}


\section{Introduction}

The problem of pattern formation is one of the most rapidly developing
branches of nonlinear science today \cite{Pel, Matk,Matk1,Matk2,reg1,reg2,reg3,reg4,reg5,reg6,reg7,OGKP,KOP99,KOP95,KOP951,KOPCOM,Thual,Joulin}.

The 2D Laplacian growth equation describes a wide range of
physical problems, for example, filtration combustion in a porous
medium, displacement of a cold liquid in a Hele-Shaw channel by
the same liquid that is heated or of a hot gas in a Hele-Shaw
channel by the same gas that is cooled, and solidification of a
solid penetrating a liquid in a channel \cite{Matk,Matk1,Matk2}.
This equation has elegant analytical solutions (\cite{NEW6} and
references inside). The subset of the solutions that have big
physical sense can be written in the form of logarithmic  poles
("Nevertheless, a considerable subclass of the purely logarithmic
solutions is well defined for all positive times and describes a
non-singular interface dynamics at zero surface tension"
\cite{NEW6}). However, such an equation can lead to the appearance
of final time singularities. To prevent these singularities and to
regularise the problem, a term containing the surface tension is
usually introduced into the equation describing Laplacian growth.
Unfortunately, in the presence of such a surface tension term,
obtaining an analytical solution in the form of poles becomes
impossible. In addition, it is usually assumed that the surface
tension explains the occurrence of an asymptotic solution in the
form of a finger with half of the channel width. This asymptotic
behaviour is also observed in experiments. In this paper, the
mathematical mechanism of the regularisation is introduced. It
makes it possible to avoid final time singularities, results in
desirable asymptotic behaviour in the form of a finger with half
of the channel width, and maintains the analytical solution in the
form of poles. Maintenance of the analytical character of the
solution is very important - it makes it possible to easily
analyse 2D Laplacian growth solutions and to qualitatively or
quantitatively explain the behaviour. The author sincerely hopes
that this paper will play the same role for the 2D Laplacian
growth equation as the paper \cite{Thual} did for the theory of
gaseous combustion of pre-mixed flames. The \cite{Thual}
analytical solution and its asymptotic behaviour have given a push
to development of the theory of gaseous combustion of pre-mixed
flames and have made it possible to qualitatively or
quantitatively explain the behaviour of a front of pre-mixed
flames \cite{OGKP,KOP99,KOP95,KOP951,KOPCOM,Joulin}.

 Matkowsky,  Aldushin \cite{Matk,Matk1,Matk2} considered planar,
uniformly propagating combustion waves driven by the filtration of
gas containing an oxidiser, which reacts with the combustible
porous medium through which it moves. These waves were typically
found to be unstable with respect to hydrodynamic perturbations
for both forward (coflow) and reverse (counterflow) filtration
combustion (FC), in which the direction of gas flow is the same as
or opposite to the direction of propagation of the combustion
wave, respectively.

The basic mechanism leading to instability is the reduction of the
resistance to flow in the region of the combustion products due to
an increase of the porosity in that region. Another destabilising
effect in forward FC is the production of gaseous products in the
reaction. In reverse FC, this effect is stabilising. In the case in
which the planar front is unstable, an alternative mode of propagation
in the form of a finger propagating with constant velocity was proposed.
The finger region occupied by the combustion products is separated
from the unburned region by a front in which chemical reactions and
heat and mass transport occur.

In the paper of Matkowsky,  Aldushin \cite{Matk,Matk1,Matk2}, it
was shown that the finger solution of the combustion problem can
be characterised as a solution of a Saffman-Taylor (ST) problem
originally formulated to describe the displacement of one fluid by
another having a smaller viscosity in a porous medium or in a
Hele-Shaw configuration. The ST problem is known to possess a
family of finger solutions, with each member characterised by its
own velocity and each occupying a different fraction of the porous
channel through which it propagates. The scalar field governing
the evolution of the interface is a harmonic function. It is
natural, then, to call the whole process $Laplacian$ $growth$.

The mathematical problem of Laplacian growth without surface tension
exhibits a family of exact analytical solutions in terms of logarithmic
poles in the complex plane.

The main problem with such a solution is existing finite time singularities.
To prevent such singularities, nonzero surface tension usually is
used (\cite{reg1,reg2,reg3,reg4,reg5}). The surface tension also
results in a well-defined asymptotic solution: only one finger with
half of the channel width. In addition, the other terms can be used
for regularisation (see \cite{reg6,reg7} and references therein).

The solution of the ST problem without surface tension is similar
to the solution for the equation describing cellular flames in the
case of combustion of gas mixtures \cite{OGKP,KOP99,KOP95,KOP951}.
Indeed, in both cases, solutions can be transformed to the set of
ordinary differential equations. This set describes the motion of
poles in the complex plane. Applying nonzero surface tension to the
ST problem destroys this elegant analytical solution.

It must be mentioned that the filtration combustion and the gaseous
combustion in pre-mixed flames are features of different physics;
the equation of 2D Laplacian growth and the equation describing the
Mihelson-Sivashinsky feature use completely different mathematics.
Moreover, whereas the equation for the Mihelson-Sivashinsky poles
involves trigonometric functions, the equation for the 2D Laplacian
growth poles involves logarithmic functions. The analogy here is not
"half-baked" but rather is deep. Indeed,
the very different and complex integro-differential equations have
a simple analytical solution in the form of poles. Moreover, even
the behaviour of these poles is similar.

Another problem is the fact that surface tension may not be introduced
for the mathematical problem considered by Saffman and Taylor involving
filtration combustion in a porous medium \cite{Matk,Matk1,Matk2}.
Here, the zone of chemical reaction and diffusion of heat and mass
shrinks to an interface separating the burned region from the unburned
region. In all these problems, there is no pressure jump at the interface,
so surface tension may not be introduced. Thus, the Saffman-Taylor
model arises not only as the limiting case of zero surface tension
in a problem in which surface tension enters the problem in a natural
way but also in other situations in which the introduction of surface
tension makes no sense. Another such problem is that of the solidification
of a solid penetrating a liquid in a channel. It is reasonable to
expect that the selection may be affected by introducing a perturbation
other than surface tension that is relevant to the specific problem
under consideration. For example, in the combustion problem, the effect
of diffusion as a perturbation might be considered. Here, the effect
of diffusion is similar to that of surface tension in the fluid displacement
problem \cite{Matk,Matk1,Matk2}.

Therefore, we need to look for a solution without introducing surface
tension using different methods for regularisation.

Criteria were proposed (\cite{Matk1} and \cite{Matk2}) to select
the correct member of the family of solutions (one finger with half
of the channel width) based on a consideration of the ST problem itself,
rather than on modifications of the problem. A modification is obtained
by adding surface tension to the model and then taking the limit of
the vanishing surface tension.

It is nice that we know now the criteria for the correct
asymptotic solution. Unfortunately, it is not clear from the
papers \textit{why}  Laplacian growth without surface tension
gives this asymptotic solution (one finger with half of the
channel width) that satisfies the identified criteria (namely,
which mathematical mechanism results in regularization of the
problem). No proof exists in \cite{Matk1,Matk2}  that the
asymptotic solution must obey these criteria. These criteria are
not derived theoretically from the motion equations, but are
invented by authors from the knowledge of the experimental
asymptotic solution.

In this paper, we introduce mathematical mechanism of
regularization, which is not based on surface tension, and get the
asymptotic solution. First of all, we can introduce a small amount
of noise to the system. (The noise can be considered a pole flux
from infinity.) Second, for regularization of the problem, we
throw out all new poles that can produce a finite time
singularity. It can be strictly proved that the asymptotic
solution for such a system is a single finger. Moreover, the
qualitative consideration demonstrates that a finger with
$\frac{1}{2}$ of the channel width is statistically stable.
Therefore, all properties of such a solution are exactly the same
as for the solution with a nonzero surface tension under numerical
noise.

The rest of the paper is organised as follows. Next, Section 2
describes asymptotic single Saffman-Taylor "finger" formation
without surface tension. Then we present arguments about
Saffman-Taylor "finger" formation with half of the channel size
(Section 3). Finally (Section 4), we provide a summary and
conclusions.

\section{Asymptotic single Saffman-Taylor  ``finger"
formation without surface tension}

\label{sec:method1}

In the absence of surface tension, the effect of which is to stabilise
the short-wavelength perturbations of the interface, the problem of
2D Laplacian growth is described as follows:

\begin{equation}
(\partial_{x}^{2}+\partial_{y}^{2})u=0\ .\label{oqz2}
\end{equation}

\begin{equation}
u\mid_{\Gamma(t)}=0\ ,\partial_{n}u\mid_{\Sigma}=1\ .\label{oqz3}
\end{equation}

\begin{equation}
v_{n}=\partial_{n}u\mid_{\Gamma(t)}\ .\label{oqz4}
\end{equation}

Here, $u(x,y;t)$ is the scalar field mentioned, $\Gamma(t)$ is the
moving interface, $\Sigma$ is a fixed external boundary, $\partial_{n}$
is a component of the gradient normal to the boundary (i.e. the normal
derivative), and $v_{n}$ is a normal component of the velocity of
the front.

Now, we introduce physical ``no-flux" boundary conditions.
This means no flux occurs across the lateral boundaries of the channel.
This requires that the moving interface orthogonally intersects the
walls of the channel. However, unlike the case of periodic boundary
conditions, the end points at the two boundaries of the channel do
not necessarily have the same vertical coordinate. Nevertheless, this
can also be considered as a periodic problem in which the period equals
\emph{twice} the width of the channel. However, only half of this
periodic strip should be considered as the physical channel, whereas
the second half is its unphysical mirror image.

Then, we introduce a time-dependent conformal map $f$ from the
lower half of a ``mathematical" plane, $\xi\equiv\zeta+i\eta$, to
the domain of the physical plane, $z\equiv x+iy$, where the
Laplace equation \ref{oqz2} is defined as
$\xi\stackrel{f}{\longrightarrow}z$. We also require that
$f(\xi,t)\approx\xi$ for $\xi\longrightarrow\zeta-i\infty$. Thus,
the function $z=f(\zeta, t)$ describes the moving interface. From
Eqs. (\ref{oqz2}), (\ref{oqz3}), and (\ref{oqz4}) for function
$f(\xi, t)$ we obtain the $Laplacian$ $Growth$ $Equation$:

\begin{equation}
Im(\frac{\partial f(\xi,t)}{\partial\xi}\frac{\overline{\partial
f(\xi,t)}}{\partial t})=1\mid_{\xi=\zeta-i0}\
,f_{\xi}\mid_{\zeta-i\infty}=1\ .\label{oqz5}
\end{equation}

Let us look for a solution of Eq. (\ref{oqz5}) in the following form:

\begin{eqnarray}
 &  & f(\xi,t)=\lambda\xi-i{\tau(t)}-i\sum_{l=1}^{N}\alpha_{l}\log(e^{i{\xi}}-e^{i{\xi_{l}}(t)}),\label{oqz6}
\end{eqnarray}

\begin{eqnarray}
 &  & z(\zeta, t)=f(\zeta, t)=\lambda\zeta-i{\tau(t)}-i\sum_{l=1}^{N}\alpha_{l}\log(e^{i{\zeta}}-e^{i{\xi_{l}}(t)}),\label{oqz6}
\end{eqnarray}

\begin{eqnarray}
\alpha=\sum_{l=1}^{N}\alpha_{l}=1-\lambda,\label{oqz66}
\end{eqnarray}

where $\tau(t)$ is some real function of time, $\alpha_{l}$ is a
complex constant, $\xi_{l}=\zeta_{l}+i\eta_{l}$ denotes the position
of the pole with the number $l$, and $N$ is the number of poles.

For our ``no-flux" boundary condition, we must add the condition
that for every pole $\xi_{l}=\zeta_{l}+i\eta_{l}$ with
$\alpha_{l}$ exists a pole $\xi_{l}=-\zeta_{l}+i\eta_{l}$ with
$\overline{\alpha_{l}}$.

Therefore, we can conclude from this condition for pairs of poles
and eq. (\ref{oqz66}) that $\lambda$ is a real constant.

We will prove below that the necessary condition for no finite time
singularities for a pole solution is
\begin{eqnarray}
-1<\lambda<1\ ,\label{oqz666}
\end{eqnarray}

Also, for the function $F(i{\xi},t)=if(\xi,t)$, for the ``no-flux"
boundary condition,
\begin{equation}
\overline{F(i{\xi},t)}=F(\overline{i\xi},t)
\end{equation}

We want to prove that the final state will be only one finger if no
finite time singularity appears during poles evolutions.

\subsection{Asymptotic behaviour of the poles in the mathematical plane}

This derivation is similar to \cite{KOP953}, but we also consider
``no-flux" boundary conditions here (in analogy with
\cite{Fei953}).

The main purpose of this chapter is to investigate the asymptotic
behaviour of the poles in the mathematical plane. We want to
demonstrate that for time $t\mapsto\infty$, all poles go to the
two boundary points for no-flux boundary conditions or to a single
point  for periodic boundary conditions.

The equation for the interface is
\begin{eqnarray}
 &  & f(\xi,t)=\lambda\xi-i{\tau(t)}-i\sum_{l=1}^{N}\alpha_{l}\log(e^{i\xi}-e^{i{\xi_{l}}(t)}),\nonumber \\
 &  & \sum_{l=1}^{N}\alpha_{l}=1-\lambda,-1<\lambda<1\ .\label{oqf1}
\end{eqnarray}

By substitution of Eq. (\ref{oqf1}) in the $Laplacian$ $Growth$
$Equation$,
\begin{equation}
Im(\frac{\partial f(\xi,t)}{\partial\xi}\frac{\overline{\partial f(\xi,t)}}{\partial t})=1\mid_{\xi=\zeta-i0}\ ,\label{oqf2}
\end{equation}

\begin{figure}[h!]
    \centering
    \includegraphics[scale=0.38]{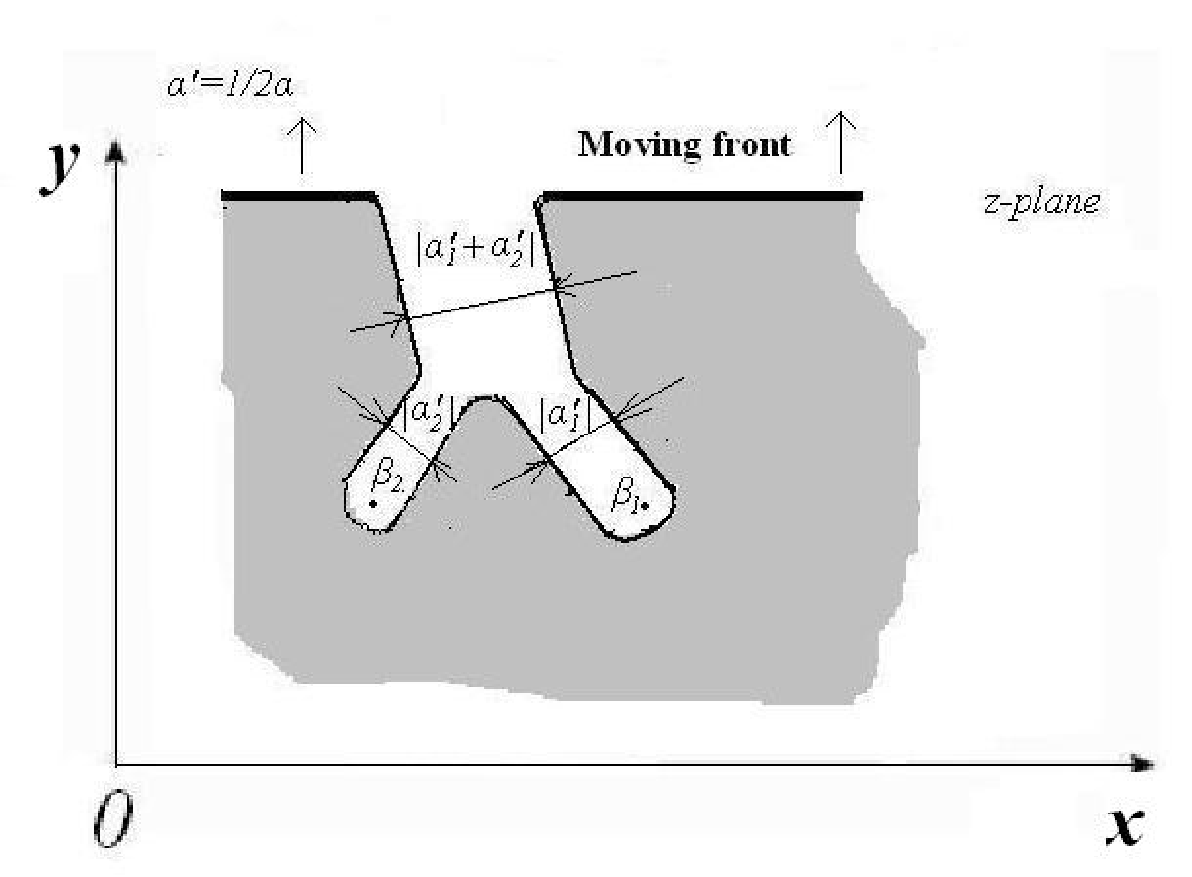}
    \caption{Geometrical interpretation of the complex constants
of motion $\alpha_{k}^{\prime}=\frac{1}{2}\alpha_{k}$ and
$\beta_{k}$; $k=1,...,N$.}
\end{figure}

we can find the equations of pole motion (Fig. 1):
\begin{equation}
\beta_{l}=\tau(t)+(1-\sum_{k=1}^{N}\overline{\alpha_{k}})\log\frac{1}{a_{l}}+\sum_{k=1}^{N}\overline{\alpha_{k}}\log(\frac{1}{a_{l}}-\overline{a_{k}})=const\label{oqf3}
\end{equation}
and
\begin{equation}
\tau=t-\frac{1}{2}\sum_{k=1}^{N}\sum_{l=1}^{N}\overline{\alpha_{k}}\alpha_{l}\log(1-\overline{a_{k}}a_{l})+C_{0}\ ,\label{oqf4}
\end{equation}
where $a_{l}=e^{i\xi_{l}}$ and $C_{0}$ is a constant.

From eqs. (\ref{oqf3}), we can find
\begin{eqnarray}
(1-\lambda)\tau-\sum_{l=1}^{N}\alpha_{l}\log a_{l}+\nonumber \\
\sum_{k=1}^{N}\sum_{l=1}^{N}\overline{\alpha_{k}}\alpha_{l}\log(1-\overline{a_{k}}a_{l})=const\ .\label{oqf6}
\end{eqnarray}
From eqs. (\ref{oqf4}) and (\ref{oqf6}), we can obtain
\begin{equation}
Im(\sum_{l=1}^{N}\alpha_{l}\log a_{l})=const\label{oqf7}
\end{equation}
and
\begin{equation}
t=(\frac{1+\lambda}{2})\tau+\frac{1}{2}Re(\sum_{l=1}^{N}\alpha_{l}\log a_{l})+C_{1}/2\ ,\label{oqf8}
\end{equation}
where $C_{1}$ and $\alpha_{l}$ are constants, $\xi_{l}(t)$ is the
position of the poles, and $a_{l}=e^{i{\xi_{l}}(t)}$.

In Appendix A, we will prove from eq.(\ref{oqf4}) that $\tau\mapsto\infty$
if $t\mapsto\infty$ and if no finite time singularity exists.

The equations of pole motion that follow from eqs. (\ref{oqf3}) are
as follows:
\begin{equation}
\tau+i\overline{\xi_{k}}+\sum_{l}\alpha_{l}\log(1-e^{i(\xi_{l}-\overline{\xi_{k}})})=const,\label{aqf9}
\end{equation}

or in a different form:
\begin{eqnarray}
\zeta_{k}+\sum_{l}(\alpha_{l}^{\prime\prime}\log\mid1-e^{i(\xi_{l}-\overline{\xi_{k}})}\mid+\nonumber \\
\alpha_{l}^{\prime}\arg(1-e^{i(\xi_{l}-\overline{\xi_{k}})}))=const,\label{oqf9}
\end{eqnarray}

\begin{eqnarray}
{\tau}+\eta_{k}+\sum_{l}(\alpha_{l}^{\prime}\log\mid1-e^{i(\xi_{l}-\overline{\xi_{k}})}\mid-\nonumber \\
\alpha_{l}^{\prime\prime}\arg(1-e^{i(\xi_{l}-\overline{\xi_{k}})}))=const,\label{oqf10}
\end{eqnarray}
where
\begin{equation}
\xi_{l}=\zeta_{l}+i\eta_{l},\eta_{l}>0\ .\label{oqf11}
\end{equation}

\begin{equation}
\alpha_{l}=\alpha_{l}^{\prime}+i\alpha_{l}^{\prime\prime}\ .\label{oqf11rr}
\end{equation}

Let us transform
\begin{eqnarray}
\arg(1-e^{i(\xi_{l}-\overline{\xi_{k}}})=\nonumber \\
\arg([1-e^{i(\zeta_{l}-\zeta_{k})}e^{-(\eta_{l}+\eta_{k})}])=\nonumber \\
\arg[1-a_{lk}e^{i\varphi_{lk}}]\label{oqf13}
\end{eqnarray}

\begin{equation}
\varphi_{lk}=\zeta_{l}-\zeta_{k},a_{lk}=e^{-(\eta_{l}+\eta_{k})}\label{oqf14}
\end{equation}

$\arg[1-a_{lk}e^{i\varphi_{lk}}]$ is a single-valued function of
$\varphi_{lk}$, i.e.,

\begin{equation}
-\frac{\pi}{2}\leq\arg[1-a_{lk}e^{i\varphi_{lk}}]\leq\frac{\pi}{2}\ .\label{oqf15}
\end{equation}

We multiply eq. (\ref{oqf10}) by $\alpha_{k}^{\prime\prime}$ and
eq. (\ref{oqf9}) by $\alpha_{k}^{\prime}$, and taking the difference,
we obtain the following equation:
\begin{eqnarray}
\alpha_{k}^{\prime}\zeta_{k}-\alpha_{k}^{\prime\prime}\tau+\nonumber \\
\sum_{l\neq k}((\alpha_{l}^{\prime\prime}\alpha_{k}^{\prime}-\alpha_{k}^{\prime\prime}\alpha_{l}^{\prime})\log\mid1-e^{i(\xi_{l}-\overline{\xi_{k}})}\mid+\nonumber \\
(\alpha_{l}^{\prime}\alpha_{k}^{\prime}+\alpha_{l}^{\prime\prime}\alpha_{k}^{\prime\prime})\arg(1-e^{i(\xi_{l}-\overline{\xi_{k}})}))=const.\label{aqf10}
\end{eqnarray}

We want to investigate the asymptotic behaviour of poles $\tau\mapsto\infty$.

We have the divergent terms $\alpha_{k}^{\prime\prime}\tau$ in
this equation. From eq. (\ref{aqf10} ), only the term
$\log\mid1-e^{i(\xi_{l}-\overline{\xi_{k}})}\mid$ can eliminate
this divergence. The necessary condition for this to occur is
$\eta_{k}\mapsto0$ for $\tau\mapsto\infty,1\leq k\leq N$.


We may assume that for $t\mapsto\infty$, $N^{\prime}$ groups of
poles exist (${N^{\prime}\leq N}$) ($\varphi_{lk}\mapsto0$ for all
members of a group). The $N^{\prime}$ is currently arbitrary and
can even be equal to $N$. $N_{l}$ is the number of poles in each
group, $1\leq l\leq N^{\prime}$.

For each group, by summation of eqs. (\ref{aqf10}) over all group
poles, we obtain
\begin{eqnarray}
\alpha_{k}^{gr\prime}\zeta_{k}^{gr}-\alpha_{k}^{gr\prime\prime}\tau+\nonumber \\
\sum_{l\neq k}((\alpha_{l}^{gr\prime\prime}\alpha_{k}^{gr\prime}-\alpha_{k}^{gr\prime\prime}\alpha_{l}^{gr\prime})\log\mid1-e^{i(\xi_{l}^{gr}-\overline{\xi_{k}^{gr}})}\mid+\nonumber \\
(\alpha_{l}^{gr\prime}\alpha_{k}^{gr\prime}+\alpha_{l}^{gr\prime\prime}\alpha_{k}^{gr\prime\prime})\arg(1-e^{i(\xi_{l}^{gr}-\overline{\xi_{k}^{gr}})}))=const,\label{aqf11}
\end{eqnarray}
where
\begin{equation}
\alpha_{l}^{gr\prime\prime}=\sum_{k}^{N_{l}}\alpha_{k}^{\prime\prime}\ ,\label{oqf50}
\end{equation}

\begin{equation}
\alpha_{l}^{gr\prime}=\sum_{k}^{N_{l}}\alpha_{k}^{\prime}\ .\label{oqf51}
\end{equation}

We have no merging between defined groups for large $\tau$, so we
investigate the motion of poles with this assumption:
\begin{equation}
\mid\zeta_{l}^{gr}-\zeta_{k}^{gr}\mid\gg\eta_{l}^{gr}+\eta_{k}^{gr},1\leq l,k\leq N\ .\label{oqf16}
\end{equation}

For $l\neq k$, $\eta_{k}^{gr}\mapsto0$, and $\varphi_{lk}^{gr}=\zeta_{l}^{gr}-\zeta_{k}^{gr}$,
we obtain
\begin{eqnarray}
\log\mid1-e^{i(\xi_{l}^{gr}-\overline{\xi_{k}^{gr}})}\mid\approx\log\mid1-e^{i(\zeta_{l}^{gr}-\zeta_{k}^{gr})}\mid=\nonumber \\
\log2+\frac{1}{2}\log\sin^{2}\frac{\varphi_{lk}^{gr}}{2}\label{oqf20}
\end{eqnarray}
and
\begin{eqnarray}
\arg(1-e^{i(\xi_{l}^{gr}-\overline{\xi_{k}^{gr}})})\approx\arg(1-e^{i(\zeta_{l}^{gr}-\zeta_{k}^{gr})})=\nonumber \\
\frac{\varphi_{lk}^{gr}}{2}+\pi n-\frac{\pi}{2}\ .\label{1qf20}
\end{eqnarray}
We choose $n$ in Eq.(\ref{1qf20}) so that Eq.(\ref{oqf15}) is correct.
Substituting these results into eqs. (\ref{aqf11}), we obtain
\begin{eqnarray}
C_{k}={\alpha^{gr\prime}{\zeta_{k}^{gr}}}-\alpha_{k}^{gr\prime\prime}\tau+\nonumber \\
\sum_{l\neq k}[({\alpha_{l}^{gr\prime\prime}\alpha_{k}^{gr\prime}}-{\alpha_{k}^{gr\prime\prime}\alpha_{l}^{gr\prime}})\log\mid\sin\frac{\varphi_{lk}^{gr}}{2}\mid\nonumber \\
+(\alpha_{l}^{gr\prime}\alpha_{k}^{gr\prime}+{\alpha_{l}^{gr\prime\prime}\alpha_{k}^{gr\prime\prime})}\frac{\varphi_{lk}^{gr}}{2}].\label{oqf30}
\end{eqnarray}

\subsection{Theorem about coalescence of the poles}

From eqs. (\ref{oqf30}), we can conclude the following:

(i) By summation of eqs. (\ref{oqf30}) (or exactly from eq. (\ref{oqf7})),
we obtain
\begin{equation}
\sum_{k}\alpha_{k}^{gr\prime}{\zeta_{k}^{gr}}=const\ .\label{oqf31}
\end{equation}

(ii) For $\mid{\varphi}_{lk}^{gr}\mid\mapsto0,2\pi$, we obtain $\log\mid\sin\frac{\varphi_{lk}^{gr}}{2}\mid\mapsto\infty$,
meaning that the poles can not pass each other;

(iii) From (ii), we conclude that $0<\mid\varphi_{lk}^{gr}\mid<2\pi$;

(iv) From (i) and (iii), $\zeta_{k}^{gr}\mapsto\infty$ is impossible;

(v) In eq.(\ref{oqf30}), we must compensate for the second divergent
term. From (iv) and (iii), we can do this only if $\alpha_{l}^{gr\prime\prime}=\sum_{k}^{N_{l}}\alpha_{k}^{\prime\prime}=0$
for all $l$.

Therefore, from eq. (\ref{oqf30}), we obtain
\begin{equation}
\sum_{k}^{N_{l}}\alpha_{k}^{\prime\prime}=0\ ,\label{oqf52}
\end{equation}

\begin{equation}
\dot{\varphi_{lk}^{gr}}=0\ ,\label{oqf53}
\end{equation}

\begin{equation}
{\varphi_{lk}^{gr}}\neq0\ ,\label{oqf54}
\end{equation}

\begin{equation}
\dot{\zeta_{k}^{gr}}=0\ .\label{oqf55}
\end{equation}

For the asymptotic motion of poles in group $N_{m}$, we obtain the
following from eqs. (\ref{oqf52}), (\ref{oqf53}), (\ref{oqf54}),
and (\ref{oqf55}), taking the leading terms in eqs. (\ref{oqf8})
and (\ref{aqf9}):
\begin{equation}
\tau=\frac{2}{\lambda+1}t\ ,\label{oqf57}
\end{equation}

\begin{equation}
0=\dot{\tau}+\sum_{l}^{N_{m}}\alpha_{l}\frac{\dot{\eta_{k}}+\dot{\eta_{l}}+i(\dot{\zeta_{k}}-\dot{\zeta_{l}})}{\eta_{k}+\eta_{l}+i(\zeta_{k}-\zeta_{l})}\ .\label{oqf56}
\end{equation}

The solution to these equations is
\begin{equation}
\eta_{k}=\eta_{k}^{0}e^{-\frac{1}{\alpha_{m}^{gr\prime}}\frac{2}{1+\lambda}t}\ ,\label{oqf58}
\end{equation}

\begin{equation}
\varphi_{lk}={\varphi_{lk}}^{0}e^{-\frac{1}{\alpha_{m}^{gr\prime}}\frac{2}{1+\lambda}t}\ ,\label{oqf59}
\end{equation}

\begin{equation}
\dot{\zeta_{k}}=0\ .\label{oqf60}
\end{equation}

Therefore, we may conclude that to eliminate the divergent term, we
need
\begin{equation}
\alpha_{l}^{gr\prime\prime}=\sum_{k}^{N_{l}}\alpha_{k}^{\prime\prime}=0,\label{orf52}
\end{equation}

\begin{equation}
\alpha_{l}^{gr\prime}(1+\lambda)>0
\end{equation}
for all $l$.


\subsection{The final result}
With the  periodic boundary condition, eq.(\ref{orf52}) is correct
for all poles, so we obtain
       $N^{\prime}=1$,
       $m=1$ and $N_m=N$.

       Therefore the unique solution is

\begin{equation}
     \eta_k=\eta_k^0e^{-{2\over (1-\lambda^2)} t} \ , \label{oqf61}
\end{equation}

\begin{equation}
            \varphi_{lk}={\varphi_{lk}}^0e^{-{2\over (1-\lambda^2)} t}
     \ , \label{oqf62}
\end{equation}

\begin{equation}
     \dot{\zeta_k}=0 \ . \label{oqf63}
\end{equation}

\begin{equation}
   1-\lambda^2>0
\end{equation}

With the no-flux boundary condition, we have a pair of poles whose
condition in eq. (\ref{orf52}) is correct, so all these pairs must
merge. Because of the symmetry of the problem, these poles can merge
only on the boundaries of the channel $\zeta=0,\pm\pi$. Therefore,
we obtain two groups of the poles on boundaries. $N^{\prime}=2$,
$m=1,2$, $N_{1}+N_{2}=N$, and $\alpha_{1}^{gr\prime}+\alpha_{2}^{gr\prime}=1-\lambda$.
(In principle, it is possible for some degenerate case of $\alpha_{l}^{gr\prime}$
values that eq. (\ref{orf52}) would be correct for some different
groups of poles. However, this is a very improbable, rare case.)

\begin{figure}[h!]
    \centering
    \includegraphics[scale=0.38]{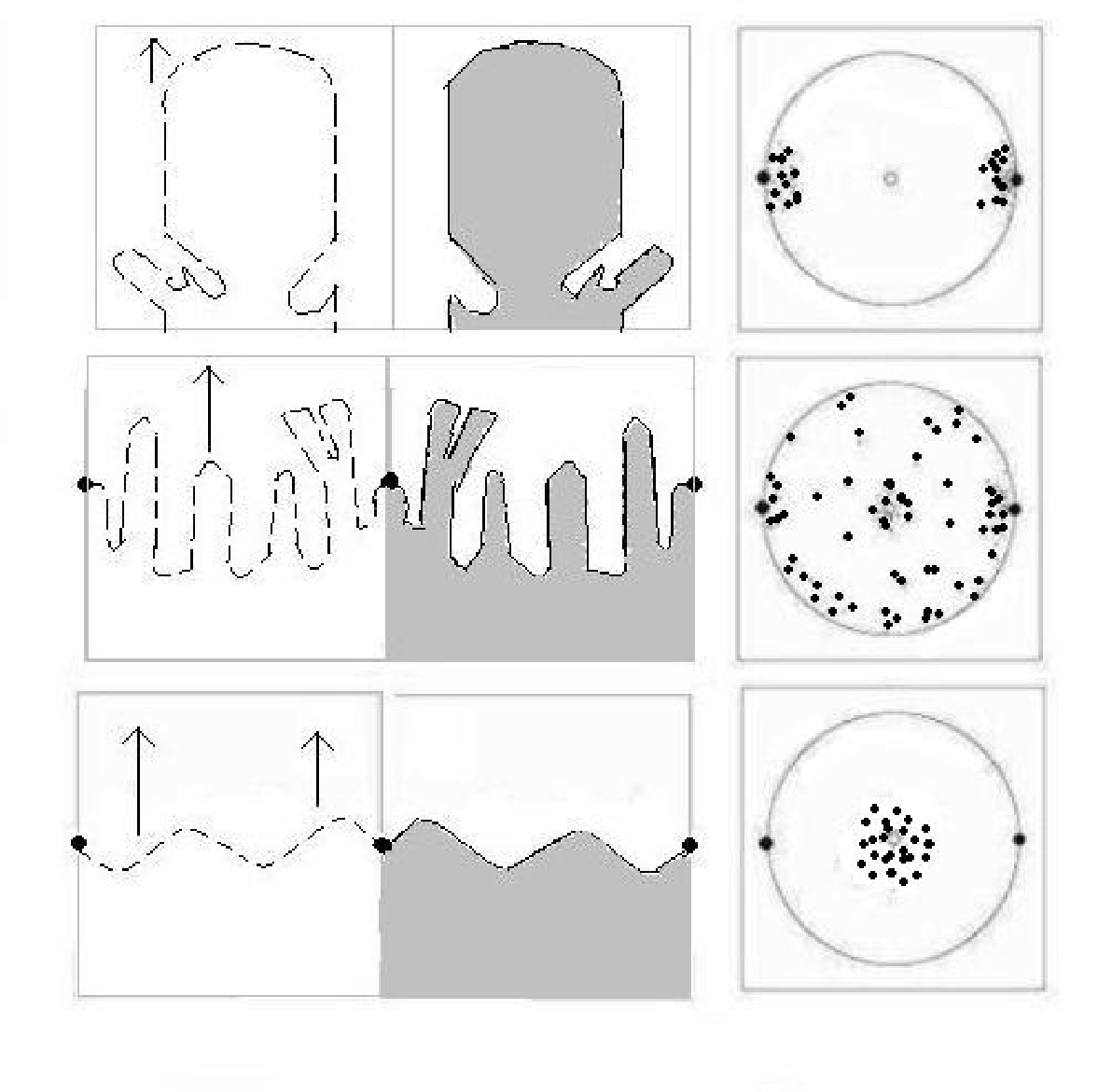}
    \caption{Three consecutive stages of fingering in the Hele-Shaw
cell: initial (left), intermediate (center), and asymptotic
(right). The physical plane $z$ is shown in the upper pictures,
while the lower pictures depict a distribution of moving poles
$a_{k}(t)$ in the unit circle $|\omega|<1$ on the mathematical
plane $\omega$. The open circle indicates the repeller,
$\omega=0$, while the solid circle indicates the attractor,
$\omega=1$, of poles whose dynamics is given by
(\ref{oqf3}-\ref{oqf4}).}
\end{figure}

Consequently, we obtain the solution (on two boundaries of the
channel Fig. 2):
\begin{equation}
\eta_{k}^{(1)}=\eta_{k}^{(1),0}e^{-\frac{1}{\alpha_{1}^{gr\prime}}\frac{2}{1+\lambda}t}\ ,\label{otf58}
\end{equation}

\begin{equation}
\varphi_{lk}^{(1)}={\varphi_{lk}}^{(1),0}e^{-\frac{1}{\alpha_{1}^{gr\prime}}\frac{2}{1+\lambda}t}\ ,\label{otf59}
\end{equation}

\begin{equation}
{\zeta_{k}^{(1)}}=0\ ;\label{otf60}
\end{equation}

\begin{equation}
\eta_{k}^{(2)}=\eta_{k}^{(2),0}e^{-\frac{1}{\alpha_{2}^{gr\prime}}\frac{2}{1+\lambda}t}\ ,\label{otf581}
\end{equation}

\begin{equation}
\varphi_{lk}^{(2)}={\varphi_{lk}}^{(2),0}e^{-\frac{1}{\alpha_{2}^{gr\prime}}\frac{2}{1+\lambda}t}\ ,\label{otf591}
\end{equation}

\begin{equation}
{\zeta_{k}^{(2)}}=\pm\pi\ ;\label{otf601}
\end{equation}

\begin{equation}
\alpha_{1}^{gr\prime}(1+\lambda)>0,\label{otf601a}
\end{equation}

\begin{equation}
\alpha_{2}^{gr\prime}(1+\lambda)>0.\label{otf601b}
\end{equation}

By summation of eqs. (\ref{otf601a}) and (\ref{otf601a}) and using
eq. \ref{oqz66}, we obtain

\begin{equation}
(1-\lambda)(1+\lambda)=1-\lambda^{2}>0.\label{otf601c}
\end{equation}

This immediately gives us the formerly formulated condition (\ref{oqz666})
for $\lambda$.

$\frac{\lambda+1}{2}=1-\frac{\alpha}{2}$ has an explicit physical
sense. It is the portion of the channel occupied by the moving liquid.
We see that for no finite time singularity and for $t\mapsto\infty$,
we obtain one finger with width $\frac{\lambda+1}{2}$.

\section{Saffman-Taylor "finger" formation with
half of the channel size}

\label{sec:method0} The case of Laplacian growth in the channel
without surface tension was considered in detail by
Mineev-Weinstein and Dawson \cite{94DMW}. In this case, the
problem has an elegant analytical solution. Moreover, they assumed
that all major effects in the case with vanishingly small surface
tension may also occur without surface tension. This would make it
possible to apply the powerful analytical methods developed for
the no surface tension case to the vanishingly small surface
tension case . However, without additional assumptions, this
hypothesis may not be accepted.

The first objection is related to finite time singularities for
some initial conditions. Actually, for overcoming this difficulty,
a regular item with surface tension was introduced. This surface
tension item results in loss of the analytical solution. However,
regularisation may be carried out much more simply - simply by
rejecting the initial conditions that result in these
singularities.

The second objection is given in work by Siegel and Tanveer
\cite{96Sie}. There, it is shown that in numerical simulations
(supported by some semi-analytical calculations with appearance
"daughter singularity") in a case with any (even vanishingly
small) surface tension, any initial thickness "finger" extends up
to $\frac{1}{2}$ the width of the channel during finite time,
witch does not depend on value of small surface tension. The
analytical solution in a case without surface tension results in a
constant thickness of the  ``finger" equal to its initial size,
which may be arbitrary. Siegel and Tanveer, however, did not take
into account the simple fact that numerical noise introduces small
perturbation to the initial condition or even during  ``finger"
growth, which is equivalent to the remote poles, and with respect
to this perturbation, the analytical solution with a constant
``finger" is unstable.

It was shown by Mineev-Weinstein \cite{M98} that similar pole
perturbations for some initial conditions, can be extended to the
Siegel and Tanveer solutions. This positive aspect of the paper
\cite{M98} was mentioned by Sarkissian and Levine in their Comment
\cite{S98} and in Reply of Mineev-Weinstein \cite{NEW4}. In
summary, it is possible to determine that to identify the results
with and without surface tension, it is necessary to introduce a
permanent source of the new remote poles: the source may be either
external noise or an infinite number of poles in an initial
condition. Which of these methods is preferred is still an open
question.

Of course, this additional noise will insert new poles resulting
in solutions, which are different from Siegel and Tanveer solution
\cite{S98}. However, these solutions appear as a result of the
noise both without surface tension and with surface tension. Thus,
introducing the noise erases difference between equation with
surface tension and without surface tension.

In the case of flame front propagation, it was shown
\cite{OGKP,KOP99,KOP95,KOP951,KOPCOM} that external noise is
necessary for an explanation of the flame front velocity increase
with the size of the system: using an infinite number of poles in
an initial condition cannot give this result. It is interesting to
know what the situation is in the channel Laplacian growth. One of
the main results of Laplacian growth in the channel with a small
surface tension is Saffman-Taylor ``finger" formation with a
thickness equal to $\frac{1}{2}$ the thickness of the channel. To
use the analytical result obtained for zero surface tension, it is
necessary to prove that formation of the  ``finger" also takes
place without surface tension.

In our teamwork with Mineev-Weinstein \cite{KOP953}, it was shown
that for a finite number of poles at almost all allowed (in the
sense of not approaching finite time singularities) initial
conditions, except for a small number of degenerate initial
conditions, there is an asymptotic solution involving a  ``finger"
with any possible thickness\marginpar{}. Note that the solutions
and asymptotic behaviour found in \cite{KOP953} for a finite
number of poles are an idealisation but have a real sense for any
finite intervals of time between the appearance of the new poles
introduced into the system by external noise or connected to an
entrance to the system of remote poles of an initial condition,
including an infinite number of such poles. The theorem proved in
\cite{KOP953} may again be applied for this final set of new and
old poles and again yields asymptotic behaviour in the form of a
``finger", but the thickness is different. Thus, introduction of a
source of new poles results only in possible drift of the
thickness of the final  ``finger" but does not change the type of
solution.

It should be mentioned that instead of periodic boundary
conditions, much more realistic ``no flux" boundary conditions may
be introduced \cite{Fei953}. (This paper repeats the result for
single finger asymptotic behaviour already proved formerly in the
papers \cite{KOP953}. See also reference 14 in \cite{M98} and
reference 20 (and correspondent text) in \cite{KOP953}). This
result forbids a stream through a wall, which inserts additional,
probably useful restrictions on the positions, number, and
parameters of new and old poles (explaining, for example, why the
sum of all complex parameters $\alpha_{i}$ for poles gives the
real value $\alpha$ for the pole solution (5) in \cite{M98}).
However, this does not have an influence on the correctness and
applicability of the results and methods proved in \cite{KOP953}.
No new qualitative results appear as a result of introducing ``no
flux" boundary conditions. For example, one finger asymptotic
behaviour is correct for the both cases.

Mineev-Weinstein \cite{M98} tries to give proof that steady
asymptotic behaviour for Laplacian growth in a channel with zero
surface tension is a single  ``finger" with a thickness equal to
$\frac{1}{2}$ the thickness of the channel, which is unequivocally
erroneous. Indeed, the method in \cite{M98} proves and
demonstrates the instability of a ``finger" with a thickness
distinct from $\frac{1}{2}$ with respect to introducing new remote
poles. However, the instability of a ``finger" with a thickness
equal to $\frac{1}{2}$ may be proved and demonstrated by the same
method.

Such instability is justified explicitly in Comments of
Casademunt, Magdaleno  and Almgren \cite{NEW1,NEW2}.  Casademunt
and Magdaleno write, that perturbation of a finger solution
considered by Mineev-Weinstein in \cite{M98} (precisely,
perturbation of the $\lambda\zeta$ term in the conformal
representation $z(\zeta,t)$ ), represents a special case of more
general perturbation:
\begin{eqnarray}
 &  & \lambda\zeta \approx (\lambda-\lambda_{0})\zeta -i\sum_{k=1}^{N}\delta_{k}\log(e^{i{\zeta}}-\epsilon_{k}),\label{ottqz6}
\end{eqnarray}
where  $\epsilon_{k}$ is small, and
$\sum_{k=1}^{N}\delta_{k}=\lambda_{0}$ .

Perturbation considered by Mineev-Weinstein  in \cite{M98}
corresponds to the case $\lambda=\lambda_{0}$:
\begin{eqnarray}
 &  & \lambda\zeta \approx  -i\sum_{k=1}^{N}\delta_{k}\log(e^{i{\zeta}}-\epsilon_{k}),\label{ottqz61}
\end{eqnarray}
 The finger with the thickness of  $\frac{1}{2}$  corresponds to the case  $\lambda=0$.
 Such finger will be stable with respect to perturbation of Mineev-Weinstein  in \cite{M98},
 but is unstable to more common perturbation of Casademunt and Magdaleno  in \cite{NEW1}.
 It contradicts with the statment of \cite{M98}, that the solution in the form of
 the finger $\frac{1}{2}$ is stable and, corespondently, is asymtotic for all other unstable solutions.
 The answer to this objection is done by Mineev-Weinstein  in Reply \cite{NEW3}.
 He demonstrate, that perturbation of Casademunt and Magdaleno  can be easily
 transformed to the perturbation of  Mineev-Weinstein  by adding a  pole in zero.
 From this, Mineev-Weinstein  makes two conclusions:

1) Perturbation of Casademunt and Magdaleno   is unstable. Really,
for the proof of instability, it is enough to prove instability
with respect to at least one perturbation. Such perturbation is
the additive pole. As perturbation of Casademunt and Magdaleno  is
unstable and can be transformed to perturbation odf
Mineev-Weinstein , Mineev-Weinstein  concludes, that it has no
sense to used perturbation of Casademunt and Magdaleno.

2)Perturbations of Casademunt and Magdaleno  is only "small"
subset of perturbations of Mineev-Weinstein  for a case of an
additive pole in zero. Mineev-Weinstein  concludes, that
perturbations of Casademunt and Magdaleno are "much rare", than
perturbations of Mineev-Weinstein  . Therefore, it has no sense to
used perturbation of Casademunt and Magdaleno.
   The first objection paradoxically works against arguments of Mineev-Weinstein .
   Indeed, if «for the proof of instability it is enough to show instability with respect to at least one perturbation».
   It means that instability of the finger $\frac{1}{2}$ with respect to
    perturbation of Casademunt and Magdaleno  is quite enough to proof instability of the finger $\frac{1}{2}$.
    Accordingly, the finger $\frac{1}{2}$ cannot be an asymptotics.
    The second objection contains a simple mathematical error.
    Really, for the set theory, the subset can be equal to the set.
    For example, squares of the natural numbers are only a subset of the natural numbers.
    However, between both sets there is an obvious  one-to-one correspondence.
    For the current case, for each perturbation of Mineev-Weinstein  resulting in the finger $\frac{1}{2}$:
\begin{eqnarray}
 &  & \lambda\zeta \approx  -i\sum_{k=1}^{N}\delta_{k}\log(e^{i{\zeta}}-\epsilon_{k}),\label{ottqz7}
\end{eqnarray}
is possible to find the correspondent perturbation of Casademunt
and Magdaleno:
\begin{eqnarray}
 &  & 0\zeta \approx (0- (-\lambda))\zeta -i \sum_{k=1}^{N}(-\delta_{k})\log(e^{i{\zeta}}-\epsilon_{k}),\label{ottqz8}
\end{eqnarray}
returning the thickness of the finger $\frac{1}{2}$ to the initial
value. Moreover, in our teamwork \cite{KOP953}, it is shown that
for a finite number of poles, any thickness ``finger" is possible
as an asymptotic solution.However, some factor  exists which can
break describe above one-to-one correspondence between
perturbation of Mineev-Weinstein resulting in the finger
$\frac{1}{2}$  and perturbation of Casademunt and Magdaleno
returning the thickness of the finger $\frac{1}{2}$  to the
initial value. Indeed, some of these perturbation result in finite
time singularities, not asymptotic finger. If these singularities
appear more frequently for perturbation of Mineev-Weinstein then
perturbation of Casademunt and Magdaleno we can get exlusive role
of the finger $\frac{1}{2}$. Exactly this situation appears in our
consideration below (see Fig. 3).

\begin{figure}[h!]
    \centering
    \includegraphics[scale=0.38]{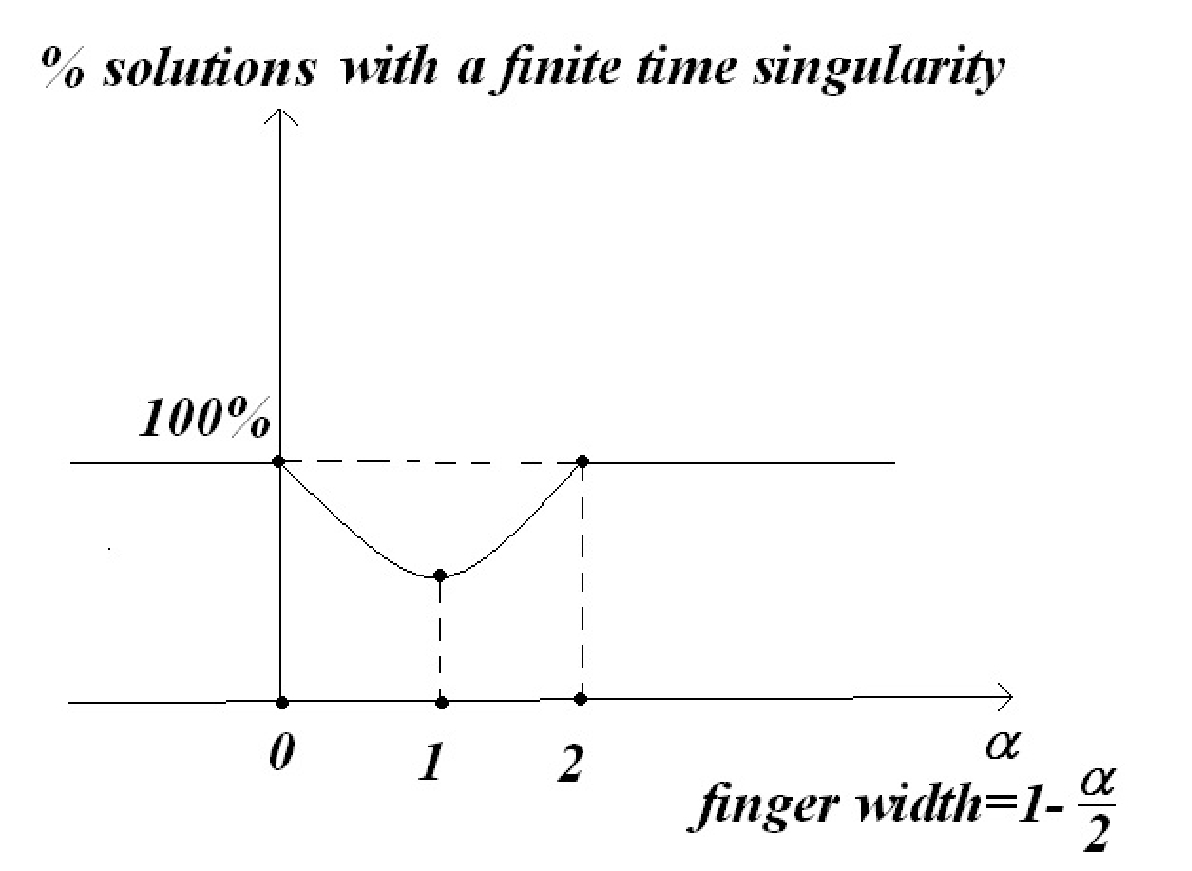}
    \caption{The width of the finger is equal to $1-\frac{\alpha}{2}$
(The channel width is assumed to be equal to 1). The graph for the
current $\alpha$ gives the percent of all possible solutions
resulting in a finite time singularity. The maximum value is equal
to 100 percent and corresponds to ${\alpha\leq0}$ or
${\alpha\geq2}$. The minimum is located at the middle point
$\alpha=1$ between $\alpha=0$ (finger width of 1) and $\alpha=2$
(finger width of 0). Therefore, at the minimum, the finger width
is $\frac{1}{2}$.}
\end{figure}

This does not mean, however, that the privileged role of a
``finger" no surface tension; it only means that the proof is not
given in \cite{M98}. Let us try to give the correct arguments
here. The general pole solution (5) in work \cite{M98} is
characterized by the real parameter $\alpha$ being the sum of the
complex parameters $\alpha_{i}$ for poles. The thickness of the
asymptotic finger is a simple function of $\alpha$: (Thickness =
$1-\frac{\alpha}{2}$). The value ($\alpha=1$) corresponds to a
thickness of $\frac{1}{2}$. As far as possible, the thickness of
the  ``finger" is between 0 and 1, and the possible $\alpha$ value
is in an interval between 0 and 2: ($0<\alpha<2$). The value
$\alpha=1$ corresponding to the finger width $\frac{1}{2}$ is
exactly in the middle of this interval. What happens to the quite
possible initial pole conditions with $\alpha$ outside of the
limits from 0 to 2? They are  ``not allowed" because of the
already identified finite time singularities \cite{KOP953}. Also,
a part of the solutions inside the interval $0<\alpha<2$ results
in similar finite time singularities.

Finding exact sufficient and necessary conditions when defining
the initial pole condition as ``not allowed", i.e., singular, is
still an open problem. How are these  ``not allowed" initial pole
conditions (to be precise, their percentage from the full number
of possible initial pole conditions corresponding to the given
real value $\alpha$) distributed inside of the interval
$0<\alpha<2$?

From reasons of continuity and symmetry with respect to $\alpha=1$
(Fig. 3), it is possible to conclude that this distribution has a
minimum at point $\alpha=1$ (thickness $\frac{1}{2}$!), the value
that is the most remote from both borders of the interval
$0<\alpha<2$, and that the distribution increases to the borders
$\alpha=2$ and 0, reaching 100 percent for all pole solutions
outside of these borders, i.e., the thickness $\frac{1}{2}$ is the
most probable because for this thickness value, the minimal
percent of initial conditions potentially capable of producing
such a thickness value is ``not allowed", i.e., results in
singularities.

A source of new poles results in drift of the finger thickness,
but this thickness drift is close to the most probable and average
size equal to $\frac{1}{2}$. A similar result is obtained in the
case of a Saffman-Taylor  ``finger" with vanishingly small surface
tension and with some external noise, which was one of the goals
of the paper.

Let's formulate shortly our conclusion. For detailed consideration
of the solution stability is necessary to consider explicitly the
noise which can be presented as a stream of poles from zero. If
such noise is not presented in computer calculations, the
numerical noise (related to terminating accuracy of evaluations)
plays a part of the explicit noise . In the presence of such
noise, it is possible to consider not asymptotic, but a stochastic
stability of the finger $\frac{1}{2}$. Such stability arises from
the regularization of the solution by rejection of poles from the
noise, which are able to lead to finite time singularities. During
such rejection for the finger $\frac{1}{2}$, the probabilities of
appearance of poles, decreasing or increasing  finger's thickness,
are identical. For a finger with a thickness in distinct from
$\frac{1}{2}$, the probability of appearance of poles,  shifting
its thickness to $\frac{1}{2}$ is more probable.

 It is interesting
(from this point of view) to consider outcomes of Kessler and
Levine \cite{NEW5}. In the paper, the case of asymmetric surface
tension converging to zero is considered. It is demonstrated, that
for very small surface tension, the asymptote is not the finger
$\frac{1}{2}$, but random noise. The Kessler and Levine conclude
about senselessness of the analysis of the asymptotic solution as
the finger $\frac{1}{2}$ for a case of the surface tension
converging to zero. However authors make the same error, as Siegel
and Tanveer in \cite{96Sie}. It is necessary to consider
explicitly, not only the surface tension, but also the noise (at
least, small numerical noise). For the small surface tension, the
big noise can lead to appearance of a new singularities before or
immediately after the disappearance of previous singularity,
smoothed by the small surface tension. It leads to the random
solution described in the paper \cite{NEW7}. I.e., for forming the
asymptotic finger $\frac{1}{2}$, it is necessary  not only a small
surface tension, but also small noise, which is not considered in
\cite{NEW5}.

It should be mentioned that these formulated arguments are only
qualitative and that a strict proof is also necessary. The first
step to this direction was made in \cite{NEW7, NEW8}.
Unfortunately, analysis of asymptotic solution for Laplacian
growth was made in \cite{NEW7} in absence of the noise. It is
physically unsensible. For asymptotic solution with noise and
regularization (by shift out unphysical and singular solutions),
results of \cite{NEW7} rather confirm the conclusions of this
paper and can be the first step for mathematical formalization of
these conclusions.

\section{Conclusions}

\label{sec:method2} The analytical pole solution for Laplacian growth
sometimes yields finite time singularities. However, an elegant solution
of this problem exists. First, we introduce a small amount of noise
to system. This noise can be considered as a pole flux from infinity.
Second, for regularisation of the problem, we throw out all new poles
that can give a finite time singularity. It can be strictly proved
that the asymptotic solution for such a system is a single finger.
Moreover, the qualitative consideration demonstrates that the finger
equal to $\frac{1}{2}$ of the channel width is statistically stable.
Therefore, all properties of such a solution are exactly the same
as those of the solution with a nonzero surface tension under numerical
noise.

Surprisingly, the flame front propagation problem (in spite of exhibiting
absolutely different physics and mathematical equations for motion)
also has analytical pole solutions and demonstrates the same qualitative
behaviour as these solutions \cite{OGKP,KOP99,KOP95,KOP951,KOPCOM}.

\newpage{}

\section{Appendix A}

We need to prove that $\tau\mapsto\infty$ if $t\mapsto\infty$ and
if no finite time singularity exists. The formula for $\tau$ is as
follows:
\begin{equation}
\tau=t+[-\frac{1}{2}\sum_{k=1}^{N}\sum_{l=1}^{N}\overline{\alpha_{k}}\alpha_{l}\log(1-\overline{a_{k}}a_{l})]+C_{0}\ ,\label{oqf65}
\end{equation}
where $\mid a_{l}\mid<1$ for all $l$.

Let us prove that the second term in this formula is greater than
zero:
\begin{eqnarray}
-\frac{1}{2}\sum_{k=1}^{N}\sum_{l=1}^{N}\overline{\alpha_{k}}\alpha_{l}\log(1-\overline{a_{k}}a_{l})=\nonumber \\
-\frac{1}{2}\sum_{k=1}^{N}\sum_{l=1}^{N}\overline{\alpha_{k}}\alpha_{l}\sum_{n=1}^{\infty}(-\frac{(\overline{a_{k}}a_{l})^{n}}{n})=\nonumber \\
\frac{1}{2}\sum_{n=1}^{\infty}\frac{1}{n}(\sum_{k=1}^{N}\overline{\alpha_{k}}(\overline{a_{k}})^{n})(\sum_{l=1}^{N}\alpha_{l}(a_{l})^{n})=\nonumber \\
\frac{1}{2}\sum_{n=1}^{\infty}\frac{1}{n}\overline{(\sum_{l=1}^{N}\alpha_{l}(a_{l})^{n})}(\sum_{l=1}^{N}\alpha_{l}(a_{l})^{n})>0
\end{eqnarray}

Therefore, the second term in eq. (\ref{oqf65}) always greater than
zero, and consequently, $\tau\mapsto\infty$ if $t\mapsto\infty$
for no finite time singularity.

\newpage{}

\noindent \textbf{Acknowledgments} We would like to thank Mark Mineev-Weinstein
for his many fruitful ideas, which were very useful for creating the
paper.

\newpage{}


\begin{thebibliography}{99}

\bibitem{Pel}
P. Pelce, Dynamics of Curved Fronts, Academic Press, Boston (1988)

\bibitem{Matk}
A.P. Aldushin, B.J. Matkowsky, Combust. Sci. Tech. 133 (1998) 293-341.

\bibitem{Matk1}
A.P. Aldushin, B.J. Matkowsky, Appl. Math. Lett. 11 (1998) 57-62.

\bibitem{Matk2}
A.P. Aldushin, B.J. Matkowsky, Phys. Fluids 11 (1999) 1287-1296.

\bibitem{reg1}
S.J. Chapman, Eur J. Appl. Math. 10 (1999) 513-534.

\bibitem{reg2}
R. Combescot, T. Dombre, V. Hakim, Y. Pomeau, Phys. Rev. Lett. 56 (1986) 2036-2039.

\bibitem{reg3}
R. Combescot, V. Hakim, T. Dombre, Y. Pomeau, A. Pumir, Phys. Rev. A 37 (1988) 1270-1283.

\bibitem{reg4}
D.C. Hong, J.S. Langer, Phys. Rev. Lett. 56 (1986) 2032-2035.

\bibitem{reg5}
B.I. Shraiman, Phys. Rev. Lett. 56 (1986) 2028-2031.

\bibitem{reg6}
S.J. Chapman, J.R. King, J. Eng. Math. 46 (2003) 1-32.

\bibitem{reg7}
S. Tanveer, J. Fluid Mech. 409 (2000) 273-308.

\bibitem  {OGKP}
Z. Olami, B. Glanti, O. Kupervasser, I. Procacccia, Phys. Rev. E 55 (1997) 2649-2663.

\bibitem{KOP99}
O. Kupervasser, Z. Olami, I. Procaccia, Phys. Rev. E 59 (1999) 2587-2593.

\bibitem{KOP95}
O. Kupervasser, Z. Olami, I. Procacccia, Phys. Rev. Lett. 76 (1996) 146-149.

\bibitem{KOP951}
B. Glanti, O. Kupervasser, Z. Olami, I. Procacccia, Phys. Rev. Lett. 80 (1998) 2477-2480.

\bibitem{KOPCOM}
O. Kupervasser, Z. Olami, Combust. Sci. Tech. 49 (2013) 141-152.

\bibitem{Thual} O. Thual, U. Frisch, M. Henon, J. Physique 46 (1985) 1485-1494.

\bibitem{Joulin} G. Joulin, J. Phys. France 50 (1989) 1069-1082.

\bibitem{94DMW} S. Ponce Dawson, M. Mineev-Weinstein, Physica 73 (1994) 373-387.

\bibitem{96Sie} M. Siegel, S. Tanveer, Phys. Rev. Lett. 76 (1996) 409-422.

\bibitem{M98} M. Mineev-Weinstein, Phys. Rev. Lett. 80 (1998) 2113-2116.

\bibitem{S98} A. Sarkissian, H. Levine, Phys. Rev. Lett. 81 (1998) 4528.

\bibitem{KOP953} M. Mineev-Weinstein, O. Kupervasser,
Formation of a Single Saffman-Taylor Finger after Fingers Competition:
An Exact Result in the Absence of Surface Tension, 82nd Statistical Mechanics Meeting, Rutgers University, 10-12 December 1999.

\bibitem{Fei953} M. Feigenbaum, I. Procaccia, B. Davidovich, J. Stat. Phys. 103 (2001) 973-1007.

\bibitem{NEW1} Casademunt J., Magdaleno F. X.,
Comment on "Selection of the Saffman-Taylor Finger Width in the
Absence of Surface Tension:An Exact Result",
 Phys. Rev. Lett. 81 (1998) 5950-5950.

\bibitem{NEW2} Almgren R.F., Comment on "Selection of the Saffman-Taylor Finger Width in the Absence
of Surface Tension:An Exact Result", Phys. Rev. Lett. 81 (1998)
5951-5951.

\bibitem{NEW3} Mineev-Weinstein M., "A Reply to the Comment
by J. Casademunt and F. X. Magdaleno, and also R.F. Almgren" ,
Phys. Rev. Lett. 81 (1998) 5952-5952


\bibitem{NEW4} Mineev-Weinstein M., "A Reply to the Comment by Armand Sarkissian and Herbert Levine" ,
Phys. Rev. Lett. 81 (1998) 4529-4529

\bibitem{NEW5} Kessler D. A., Levine H., "Microscopic Selection of Fluid Fingering Patterns",
Phys. Rev. Lett. 86 (2001) 4532-4535

\bibitem{NEW6} Ar. Abanova, M. Mineev-Weinsteinb, A. Zabrodinc, "Multi-cut solutions
of Laplacian growth", Physica D: Nonlinear Phenomena 238(17)
(2009) 1787-1796

\bibitem{NEW7} E. Paune, F.X. Magdaleno, and J. "Casademunt Dynamical Systems approach to Saffman-Taylor fingering. A Dynamical Solvability
Scenario", Physical Review E  65(5) (2002) 056213

\bibitem{NEW8} Mark Mineev-Weinstein, Gary D. Doolen, John E. Pearson,
Silvina Ponce Dawson "Formation and Pinch-off of Viscous Droplets
in the Absence of Surface Tension: an Exact Result",
http://arxiv.org/abs/patt-sol/9912006



\end{thebibliography}
\end{document}